\documentclass[superscriptaddress,twocolumn,showpacs,amsmath,amssymb,prl]{revtex4}
\usepackage[dvips]{graphicx}
\usepackage{psfrag}

\begin{document}

\title{\bf Macroscopic superpositions of superfluid flows}
\label{sec-cat}
\author{David W. Hallwood and Keith Burnett}
\affiliation{Clarendon Laboratory, University of Oxford, Parks Road, Oxford OX1
3PU, United Kingdom} 
\author{Jacob Dunningham}
\affiliation{School of Physics and Astronomy, University of Leeds, Leeds LS2 9JT,
United Kingdom} 

\begin{abstract}
We present a scheme for creating macroscopic superpositions of the direction of superfluid flow around a loop. Using the Bose-Hubbard model we study an array of Bose-Einstein condensates trapped in optical potentials and coupled to one another to form a ring. By rotating the ring so that each particle acquires on average half a quantum of superfluid flow, it is possible to create a multiparticle superposition of all the particles rotating and all the particles stationary. Under certain conditions it is possible to scale up the number of particles to form a macroscopic superposition. The simplicity of the model has allowed us to study macroscopic superpositions at an atomic level for different variables. Here we concentrate on the tunnelling strength between the potentials. Further investigation remains important, because it could lead us to making an ultra-precise quantum-limited gyroscope. 
\end{abstract}


\maketitle

Superpositions are one of the defining differences between classical and quantum mechanics. To test whether quantum mechanics can describe the macroscopic world, which would normally be described classically, we will look for superpositions in larger systems. Multiparticle superposition states have been observed in a number of systems including photons \cite{Mitchell2004a}, C${}_{60}$ molecules \cite{Arndt1999a}, and the internal state of four ${}^{9}$Be${}^{+}$ ions \cite{Sackett2000a}. Experimental signatures of larger scale quantum phenomenon were shown when Rouse \emph{et al.}~\cite{rouse_95} observed resonant tunnelling between two  macroscopically distinct states in a superconducting quantum interference device (SQUID). The observed tunnelling was between states of different flux or opposite currents flowing around a loop. Macroscopic systems consist of approximately $10^{10}$ particles or have a macroscopic measurable quantity associated with them. The currents measured in the SQUID consisted of approximately $10^9$ Cooper pairs and produce a measurable magnetic flux, meaning tunnelling between two macroscopically distinct states had been achieved. Similar systems have also been used to show cat states can be made~\cite{friedman_00, wal_00}.  

Bose-Einstein condensates (BECs) are a promising system for realising similar results. They are composed of $10^3 - 10^7$ atoms with a high proportion in the same quantum state and are sufficiently cold to undergo a quantum phase transition from superfluid to Mott insulator~\cite{greiner_02}. They also have significant advantages over SQUIDs since they are highly controllable: the coupling between condensates and the strength of the interactions between atoms can be tuned over many orders of magnitude, there are few imperfections and near perfect lattices can be created. This enables us to develop a simple model to investigate macroscopic quantum effects~\cite{jaksch_98}. SQUIDS have the advantage that a precise magnetic field can be applied to the system, which produces an easily controllable phase around the loop. This maybe more difficult for BECs. 

There have already been a number of theoretical proposals for producing cat states with BECs in a range of different set-ups~\cite{cirac}. In this paper, we present a scheme for producing a multiparticle superposition of different superfluid flow states in a ring of coupled BECs. It enables us to study quantum effects on an atomic level using the Bose-Hubbard model~\cite{jaksch_98} rather than the macroscopic approach used to describe the SQUID~\cite{friedman_00,wal_00}. We show that the macroscopic superposition can be made for larger numbers of atoms by changing the tunnelling between the sites. Cat states are important because it provides a direct manifestation of quantum mechanics at the macroscopic level in a new system. As discussed by Leggett \cite{leggett_02}, such states are important for testing the limits of validity of quantum mechanics. Once the system is understood it maybe possible to design a device that can make quantum-limited measurements of angular momentum or, equivalently, ultra-precise gyroscopes. 

Our system consists of condensed atoms trapped in an optical potential of three sites in a ring configuration. Each site is coupled to its neighbours by quantum mechanical tunnelling through the potential barriers separating them. By Larmor's theorem \cite{rosenfeld}, the analogue of applying an external flux to create a superposition in the SQUID experiments is to rotate the ring, which is equivalent to applying a phase around the ring. The same effect could be achieved by rotating the optical potential directly, producing a flow of atoms round the loop by using Bragg scattering to imprint phases on the lattice sites~\cite{saba_05, shin_05}, or producing an effective magnetic field using two resonant laser beams~\cite{jaksch_03,juzeliunas_05}. 

We can write the spatially dependent wave function for the condensed atoms as $\psi_0 (\vec{x})\! =\!e^{i \Phi (\vec{x})} |\psi_0 (\vec{x})|$, where $\Phi (\vec{x})$ is the phase of the condensate at position $x$. When the phase is not constant throughout the condensate there is a velocity field associated with it, $\vec{v}(\vec{x}) = (\hbar / m) \vec{\nabla} \Phi (\vec{x})$. The phase of the condensate must be uniquely defined at all points round the loop. For a linearly varying phase, corresponding to a flow state, this restricts the phase variation around the loop to integer multiples of $2 \pi$. In this paper we shall be looking at situations beyond the mean field theory where the atoms are in a superposition of flow states. These are found by full diagonalisation of the Bose-Hubbard Hamiltonian.

The applied phase between the sites can be incorporated into the Bose-Hubbard Hamiltonian \cite{jaksch_98} by including appropriate phase factors in the coupling terms. This gives the `twisted' Hamiltonian~\cite{rey_03}, 
\begin{eqnarray}
H \!\! &=& \!\! -J[ e^{i\phi}\left( a^\dagger b + b^\dagger c + c^\dagger a\right) + e^{-i\phi}\left(b^\dagger a + c^\dagger b + a^\dagger c \right)] \nonumber\\
&& + U ( {a^\dagger}^2 a^2 + {b^\dagger}^2 b^2 + {c^\dagger}^2 c^2),
\label{eq:ham_num}
\end{eqnarray}
where $a$, $b$ and $c$ are the annihilation operators of atoms in the three sites, $U$ is the on-site interatomic interaction strength and $J$ is the tunnelling strength between adjacent sites. The phase factors $e^{\pm i \phi}$ in the coupling terms are known as Peierl's phase factors. We note that $\phi$ does not have to obey the phase matching condition because it represents the sites and not the condensate, and it is related to the angular momentum of the sites, $L$, by $\phi = 2 \pi L /3 \hbar$.

It is convenient to consider a new orthogonal basis of operators $\{ \alpha, \beta, \gamma \}$ that incorporates the phase matching condition. One way to write this is in the quasi-momentum basis, or flow basis,  
\begin{eqnarray}
&&\alpha = (a + b + c) / \sqrt{3},\nonumber\\ 
&&\beta = (a + b e^{i 2 \pi / 3} + c e^{i 4 \pi / 3}) / \sqrt{3},\nonumber \\
&&\gamma = (a + b e^{- i 2 \pi / 3} + c e^{- i 4 \pi / 3}) / \sqrt{3}.
\label{eq:quasimodes}
\end{eqnarray}
The new basis respectively correspond to annihilation of an atom with zero flow, one quantum of clockwise flow and one quantum of anticlockwise flow. They follow the usual commutation relations and the quasi-momentum conservation rules for a periodic potential. Throughout this paper we use the convention that a positive phase variation corresponds to clockwise flow. Using this we can rewrite the `twisted' Hamiltonian in flow representation,
\begin{eqnarray}
H \!\! &=& \!\! - J \{ (2 \alpha^{\dagger} \alpha - \beta^{\dagger} \beta - \gamma^{\dagger} \gamma) \cos\phi + \sqrt{3}(\beta^{\dagger} \beta - \gamma^{\dagger} \gamma) \sin\phi \} \nonumber \\
&&\!\!+ \frac{U}{3} \{ \alpha^{\dagger}{}^2 \alpha^2 + \beta^{\dagger}{}^2 \beta^2 + \gamma^{\dagger}{}^2 \gamma^2 + 4(\alpha^{\dagger} \alpha \beta^{\dagger} \beta + \alpha^{\dagger} \alpha \gamma^{\dagger} \gamma  \nonumber \\
&&\!\!+ \beta^{\dagger} \beta\gamma^{\dagger} \gamma ) + 2( \alpha^2 \beta^{\dagger} \gamma^{\dagger} + \beta^2 \alpha^{\dagger} \gamma^{\dagger} + \gamma^2 \alpha^{\dagger} \beta^{\dagger} + h.c. \}. \label{eq:ham_mom}
\end{eqnarray}
From Eq.~(\ref{eq:ham_mom}) we see that the eigenstates of the system, when $U/J \! \ll \! 1$, are just flow states. Even in this limit Eq.~\ref{eq:ham_mom} is still valid, because the atoms still tunnel through the barriers~\cite{jaksch_98}. For $U/J \! \approx \! 1$, there is coupling between the different flow states and when $U/J \! \gg \! 1$ the system is in the Mott regime where each site acquires exactly the same number of atoms. 

Our procedure for creating cat states is as follows. Condensed atoms are trapped in the optical ring described above and the tunnelling strength between the sites is adiabatically reduced by increasing the intensity of the trapping light. This results in squeezing of the number of atoms at each site until eventually a Mott transition takes place~\cite{greiner_02}. This makes the system very stable to flow state excitations when rotated as can be seen by the absence of any phase dependence in Eq.~\ref{eq:ham_mom} when $J\!=\!0$. Next, the loop is rotated at a rate corresponding to half a quantum of flow per particle and, while it continues to be rotated, the tunnelling through the barriers is adiabatically increased. As long as the system evolves adiabatically the system will stay in the ground state, so the experiment can be considered at zero temperature. We claim this procedure creates a cat state of the form,
\begin{equation}
(|N,0,0\rangle + |0,N,0\rangle)/\sqrt{2}, \label{eq:cat}
\end{equation}
 where the three terms in the ket represent the number of atoms in the $\alpha$, $\beta$ and $\gamma$ flow states respectively. 

\begin{figure}
\psfrag{b}{}
\psfrag{p}[bc]{$\phi$ ($2 \pi /3$)}
\psfrag{e}[lc]{Eigenvalues}
\includegraphics[width=7.2cm]{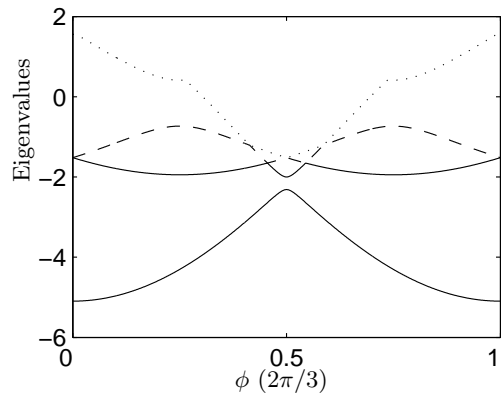}
\caption{\small The 4 lowest energy levels of the system for three atoms are shown as a function of phase, $\phi$, between two sites, in a ring of three, is varied, and $U/J\!=\!0.5$. A clear anti-crossing between the lowest two energy levels is seen.}
\label{fig:energy}
\end{figure}

Once the final state has been produced we need to show it is a cat state. One signature of an entangled state is an anti-crossing of the two lowest energy levels. In the SQUID experiment of Friedman \emph{et al.} \cite{friedman_00}, an anti-crossing of two levels was observed. This signature can be understood in the context of work on entanglement witnesses~\cite{dowling_04}. The lowest possible energy for separable states can be found using a variational approach and gives a crossing of the two lowest energy levels when the ring is rotated at a speed corresponding to half a quantum of flow per particle. This is because the zero flow state and the rotating flow state both have the same flow speed relative to the potential and therefore have the same energy. If the ground state has a lower energy than the lowest possible for separable states, we are forced to conclude that the state is entangled.

The energy eigenvalues of the resultant states are plotted against phase between two sites in Fig.~\ref{fig:energy} for three atoms, with final values of $U/J\!=\!0.5$. The probability of being in flow states $| N_{\alpha}, N_{\beta}, N_{\gamma} \rangle$ was calculated, where $N_{\alpha}$, $N_{\beta}$ and $N_{\gamma}$ are the number of atoms in the $\alpha$, $\beta$ and $\gamma$ flow states respectively. At $\phi\!=\!0$ the ground state has a probability of $0.99$ of being in state $|3,0,0 \rangle$. The deviation from $1$ is due to interactions between the particles. For increasing values of phase there is only a slight drop in the probability of this state until $\phi\! \approx\! \pi/3$. At this phase the number of atoms in the system becomes important. For commensurate numbers of atoms (i.e. the ratio of atoms to lattice sites is an integer) there is an anti-crossing of the two lowest lying energy levels, which is seen in Fig.~\ref{fig:energy}. We have shown that the energy gap persists over all values of $J$ when $U\!\neq \!0$. This means that it is always possible to evolve the system sufficiently slowly to remain in the ground state. For $U\!=\!0$ the anticrossing is not observed and the ground state becomes degenerate with other flow states at $\phi\!=\! \pi/3$. For example, if the system contained three atoms there would be a 4 fold degeneracy of the $|3,0,0\rangle$, $|0,3,0\rangle$, $|2,1,0\rangle$ and $|1,2,0\rangle$ flow states. For systems with interactions, as in the case shown in Fig.~\ref{fig:energy}, the energy of states 
with atoms of differing flow become raised. This allows a metastable superfluid flow to form in the ground state, which is needed to create a cat state~\cite{leggett_00}. In the case of Fig.~\ref{fig:energy}, the probability that the ground state is in the superposition $(|3,0,0\rangle + | 0,3,0\rangle )/\sqrt{2}$ at $\phi\!=\! \pi/3$ is $0.93$ and the exited state is in the superposition $(| 3,0,0\rangle - |0,3,0\rangle )/\sqrt{2}$ is close to $1$. This shows a good cat state has been formed. If the phase is increased further the ground state becomes a state close to $|0,3,0 \rangle$. Non-commensurate numbers produce no anti-crossing and there is just a rapid change of the ground state from the state close to $| N,0,0 \rangle$ to the state close to $| 0,N,0 \rangle$. We shall see shortly how it is possible to overcome this difference between the commensurate and non-commensurate cases. The other two lines in Fig.~\ref{fig:energy} are states close to $| 2,1,0 \rangle$ and $| 1,2,0 \rangle$. Similar results for larger numbers have been obtained, but to get similar probabilities of superpositions $U$ must be reduced to avoid significant amplitudes in higher energy states. For 30 atoms to have the ground state with a probability greater than 0.9 of being in a superposition $U/J$ must be less than 0.1. 

There is already a substantial amount of work done on similar systems at the mean field level. Phenomena such as `swallow tails' can be observed in the energy levels in figures similar to Fig.~\ref{fig:energy}~\cite{machholm_03,mueller_02,wu_02}. This is not observed in our system for any chosen parameters and suggests a possible limitation of the mean-field approach. Direct comparisons with mean field should await a fully correlated calculation for larger systems.


\begin{figure}[t]
\psfrag{a}{(a)}
\psfrag{b}{(b)}
\psfrag{NC}{$N_\gamma$}
\psfrag{NB}{$N_\beta$}
\psfrag{prob}{Probability}
\includegraphics[width=7cm]{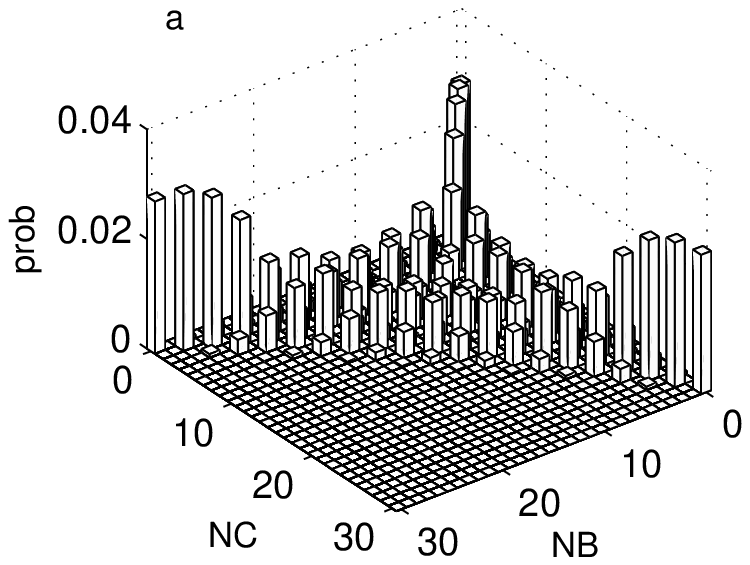}
\includegraphics[width=7cm]{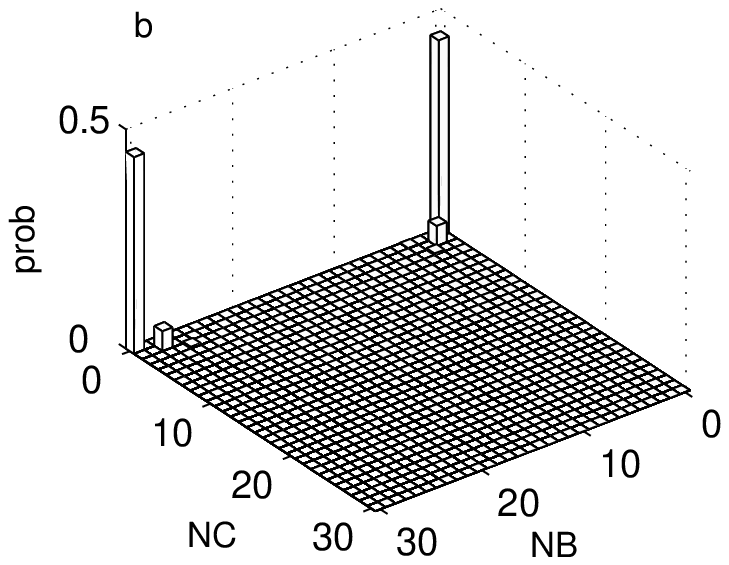}
\caption{\small (a) shows the ground state of the `twisted' Hamiltonian for 30 atoms, where $U/J=1000$ and $\phi = \pi / 3$. (b) shows the ground state for 30 atoms, where $U/J=0.1$ and $\phi = \pi / 3$. The two large peaks demonstrate a cat state. Each bar represents the probability of finding the ground state in flow state $| N-N_{\beta}-N_{\gamma}, N_{\beta},N_{\gamma}  \rangle$, where $N$ is the total number of atoms.} 
\label{fig:MomRep}
\end{figure}

The flow distribution for 30 atoms has been plotted in Fig.~\ref{fig:MomRep} (a) in the Mott regime with $\phi\!=\!\pi/3$. The distribution equally favours the states $|30,0,0\rangle$, $|0,30,0 \rangle$ and $|0,0,30\rangle$ and has a preference of atoms being in the same state, which is shown by the largest peaks being in the corners. We will show how this state can be used to create a cat. For non-commensurate number of atoms the probability distribution favours atoms being in $|N,0,0\rangle$ (or $|0,N,0 \rangle$ as there is no preference at $\phi\!=\!\pi/3$), so only peaks at one corner and is zero in the others. This state is not a suitable starting point for creating a cat by the process described here.

When the tunnelling is increased, so the ratio $U/J\! <\! 0.1$, for the commensurate case, the system is allowed to go into states $| N,0,0 \rangle$ and $| 0,N,0 \rangle$, because there was an amplitude in both these states, when the tunnelling was weak. This gives rise to the superposition shown in Fig,~\ref{fig:MomRep}~(b), where a significant proportion of the amplitude takes the form given by Eq.~\ref{eq:cat}. However, for the non-commensurate case, when $U/J\! <\!0.1$, there is no initial probability of being in $|0,N,0\rangle$ (or $|N,0,0\rangle$), so the system is forced into $|N,0,0\rangle$ (or $|0,N,0\rangle$) and no cat state is formed. A graph would show only one bar at the point representing $|N,0,0\rangle$ (or $|0,N,0\rangle$) of height close to $1$.  

\begin{figure}[t]
\psfrag{J}[cb]{$J/U$}
\psfrag{probability}[cl]{Probability}
\includegraphics[width=6cm]{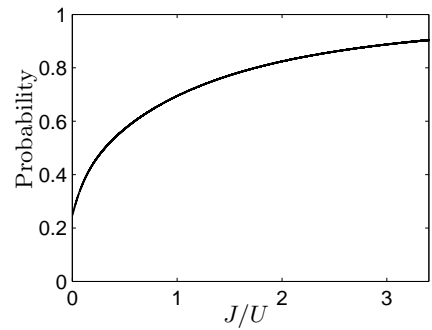}
\caption{\small Shows the probability that the ground state is in the flow superposition $(|N,0,0 \rangle + |0,N,0 \rangle )/\sqrt{2}$ as a function of $J/U$ for a total of six atoms.}
\label{fig:adiab}
\end{figure}

A simple way of overcoming the difference between the results for commensurate and non-commensurate numbers of atoms is to have slightly different tunnelling between the sites. As you can see from Eq.~\ref{eq:ham_mom} the coupling between different flow states is due to the on site interaction term. All terms in Eq.~\ref{eq:ham_mom} conserve flow (or quasi-momentum) and only $| N,0,0 \rangle$ and $|0,N,0 \rangle$ have a coupling path through intermediate flow states for commensurate numbers. For example, the state $| 3,0,0 \rangle$ can couple to $|0,3,0 \rangle$ through $| 1,1,1 \rangle$ by applying $\alpha^2 \beta^{\dagger} \gamma^{\dagger}$ then $\alpha {\beta^{\dagger}}^{2} \gamma$. Without a coupling path there is no way one flow state can evolve into another flow state, so no superposition is possible between these states. With different tunnelling between the sites, coupling between flow states where only one atom changes momentum is allowed, so $| N,0,0 \rangle$ and $|0,N,0 \rangle$ can now couple through intermediate flow states. For example, the state $| 2,0,0 \rangle$ can couple to $|0,2,0 \rangle$ through $| 1,1,0 \rangle$, which was previously not allowed. The strength of the coupling depends on, among other things, the tunnelling strength, which ultimately depends on the number of atoms. In an experimental realisation such imperfections will inevitably be present. Now when the tunnelling strength is increased, the flow distributions peak at $| N,0,0 \rangle$ and $|0,N,0 \rangle$, so a superposition similar to that shown in Fig.~\ref{fig:MomRep}~(b) is observed. The ground and excited states are close to the superposition states $| N,0,0 \rangle +e^{i \theta} |0,N,0 \rangle$ and $| N,0,0 \rangle +e^{i (\theta +\pi)} | 0,N,0 \rangle$ respectively, where the phase factor depends on the number of atoms in the system. The uneven tunnelling, therefore, aids greatly the cat making process, because there will be little distinction between commensurate and non-commensurate numbers of atoms in realistic situations and we should create cats for all numbers. A dynamical simulation of the increasing of the tunnelling strength, while the loop is rotating at a rate corresponding to half a quantum of flow, is shown in Fig.~\ref{fig:adiab} for six atoms. 

The effect of number and tunnelling strengths can be interpreted in a more intuitive way. For $U \gg J$, and equal values of $J$ around the ring, the ground state of the system is highly number squeezed. For the commensurate case, there is no net flow due to the high energy cost for doing so. This is why Fig.~\ref{fig:MomRep} (a) is symmetric. In the non-commensurate case, there are one or two extra atoms that can freely tunnel through the barriers. These particles have a net momentum and so give rise to a skewed momentum distribution.  For example, if there are 31 atoms, the system can be written as,
\begin{equation}
\overbrace{(a^{\dagger}b^{\dagger}c^{\dagger})^{10}}^{\mbox{Mott}}
\overbrace{(a^{\dagger}+e^{i\theta}b^{\dagger}+
e^{2i\theta}c^{\dagger})}^{\mbox{Superfluid}} | 0, 0, 0,\rangle, 
\label{eq:skewedmott}
\end{equation}
where the first term creates the 30 atoms in the Mott state and the second term creates the one superfluid atom that freely tunnels through the barriers. The value of $\theta$ determines the momentum state of the extra particle, which must satisfy phase matching.

When the tunnelling strengths are unequal the distribution for commensurate atoms is hardly changed. However, for the non-commensurate case, the situation changes dramatically. The atoms not trapped by the potential see a potential barrier and can either tunnel through it or be reflected from it. The condition of phase matching means that both these outcomes are both equally favourable when the phase across each pair of sites is $\phi \approx \pi/3$ and so a superposition results. For example, for 31 atoms, the state is written,
\begin{equation}
(a^{\dagger}b^{\dagger}c^{\dagger})^{10} (\alpha^{\dagger} + e^{i
\varphi}  
\beta^{\dagger}) |0,0,0\rangle, 
\end{equation}
where $\varphi$ is some phase. As the potential barriers are lowered more atoms become free and also see the barrier, so contribute to the superposition. 

There are several factors that determine how good a cat state can be produce and how accurately we need to control the rate of rotation of the ring to produce a cat. A discussion of how the tunnelling strength and number of atoms effects the cat has been presented in this paper. Future work will look at how the tunnelling strength and the number of atoms effects the required accuracy of the phase to produce a good superposition and whether long range interactions will produce more stable cats. The system only considered three points that atoms could occupy, which is very different to a SQUID that has one or several regions that are superfluid. It maybe interesting to develop a BEC system that is similar to this, because it could prove to be more stable than the three site system. Measuring whether a cat state has been formed cannot be done directly, however, the anti-crossing could be experimentally measured to give evidence of a cat using spectroscopic techniques. This is analogous to the read-out employed in the SQUID experiments and a detailed study of this will form the subject of future work.

In conclusion, we have demonstrated a scheme for creating macroscopic superposition states in the direction of the superfluid flow of Bose-Einstein condensates around a loop. This straightforward scheme relies on being able to raise, lower, and rotate an optical potential. All these techniques are within reach of current experiments. We have seen that if the tunnelling between each pair of sites is exactly equal, cats states are created only for commensurate numbers of particles. However, any slight differences in these tunnelling strengths, as we would expect to arise naturally in an experiment, means that cats can be created for any number of particles. Furthermore, the appearance of an anti-crossing between the two lowest energy levels provides a clear experimentally accessible signature that an entangled state has been formed. The simplicity of the model has allowed us to study cat states at an atomic level and future work will allow us to investigate why cat states are hard to make. This system is of great interest as it may have significant technological applications especially in the field of metrology and give new insight into quantum mechanics.


This work was supported by the United Kingdom EPSRC and by the Royal Society and Wolfson Foundation. 


\end{document}